\documentclass[conference]{IEEEtran}
\usepackage{amsthm}

\ifCLASSINFOpdf
  \usepackage[pdftex]{graphicx}
\else
\fi

\newtheorem*{definition-non}{Definition}

\begin{document}
\title{Live Visualization of GUI Application Code Coverage with GUITracer}

\author{\IEEEauthorblockA{Arthur-Jozsef Molnar \\ Faculty of Mathematics and Computer Science\\
University of Babes-Bolyai\\
Cluj-Napoca, Romania\\
arthur@cs.ubbcluj.ro}}

\maketitle

\begin{abstract}
The present paper introduces the initial implementation of a software exploration tool targeting graphical user interface (GUI) driven applications. GUITracer facilitates the comprehension of GUI-driven applications by starting from their most conspicuous artefact - the user interface itself. The current implementation of the tool can be used with any Java-based target application that employs one of the AWT, Swing or SWT toolkits. The tool transparently instruments the target application and provides real time information about the GUI events fired. For each event, call relations within the application are displayed at method, class or package level, together with detailed coverage information. The tool facilitates feature location, program comprehension as well as GUI test creation by revealing the link between the application's GUI and its underlying code. As such, GUITracer is intended for software practitioners developing or maintaining GUI-driven applications. We believe our tool to be especially useful for entry-level practitioners as well as students seeking to understand complex GUI-driven software systems. The present paper details the rationale as well as the technical implementation of the tool. As a proof-of-concept implementation, we also discuss further development that can lead to our tool's integration into a software development workflow.
\end{abstract}

\IEEEpeerreviewmaketitle

\section{Introduction}
Software tools can help practitioners in virtually all activities undertaken during the life of software starting from requirements analysis to development, program comprehension as well as test case design and execution. When studying how complex IDE's such as Eclipse evolve \cite{12,13}, we observe that newer versions ship with increasingly complex tools for aiding professionals build higher quality software faster. Modern IDE's feature tools for working with artefacts such as UML, code generation and navigation as well as supporting many common development tasks. However, we find that in most cases these tools are centred on the executable representation of the program, namely its source code and associated artefacts, limiting provided functionalities to those directly related to source code. 

Our goal is to leverage the latest results from research and industry in order to build new and useful tools for practitioners working on GUI-driven software. Our choice of this field is motivated by the fact that the GUI is the most pervasive paradigm for human-computer interaction, employed by many desktop and mobile applications. In addition, according to \cite{20}, in many cases GUI related code takes up to 50\% of application code, making it an even more compelling target. The role of tooling is already established in the literature. The authors of \cite{15} conducted a survey covering over 1400 software professionals who were inquired about their strategies, tools and the problems they encountered when comprehending software. The most significant findings show that most developers employ a white-box strategy for program comprehension. They interact with the application GUI to locate corresponding event handlers in code and in many cases use the IDE in combination with more specialized tools. Of particular note is the finding that "industry developers do not use dedicated program comprehension tools developed by the research community" \cite{15}. The purpose of our work is to provide innovative open-source tools that practitioners have a real need for and that can be used both within the academia as well as industry. After our previously developed JETracer framework \cite{10,16}, the GUITracer tool serves as the next natural step of this strategy.

This paper is structured as follows: the next section introduces the theoretical framework GUITracer is based on, while Section 3 discusses the implementation and features of the tool. The following sections discuss related work and present our conclusions as well as future work.

\section{Prerequisites}
We start from the GUI's characterization as a "hierarchical, graphical front-end to a software system that accepts as input user-generated and system-generated events" \cite{20}. As provided by Memon in \cite{20}, interaction with a GUI application can be modelled as a sequence of events. Industry studies such as \cite{15} show that practitioners approach program comprehension and feature localization tasks at the GUI level. Thus, we consider it beneficial to develop software tools that support this approach. Our goal for GUITracer is to provide a navigable relation between the target application's GUI and its underlying code. We achieve this by providing information regarding how the GUI and the underlying code are related as well as the source code that actually runs once a GUI event is fired. The following sections introduce some theoretical notions that are used within our tool as well as our previously developed JETracer framework, on which our tool is based.

\subsection{Code Coverage for GUI Events}
While the literature abounds with code coverage related techniques and tools, we find far less work focused on metrics tailored for GUI-driven applications. In this section we propose several metrics that measure the relation between GUI events and the application source code that handles them. Our approach combines established coverage criteria with Memon's event-sequence coverage defined in \cite{169}. To this we add knowledge gained from call graphs built using static analysis, which improve the code coverage picture. The first step in our effort is to define a GUI event's call graph:

\begin{definition-non}
{\bf Event call graph}. Given a GUI event $e$, we define its event call graph as a subgraph of the application's statically computed call graph that consists of event $e$'s application-level event handlers and all application methods reachable from them.
\end{definition-non}

The event call graph provides information regarding which methods might be called when handling the event, as well as call relations between them. In practice, this is computed using static analysis once the event's handlers are known. However, GUI events are not fired in isolation, but are part of an event sequence \cite{169}. Each fired event contributes to improved code coverage. We take this into account when we define an event's call graph coverage:

\begin{definition-non}
{\bf Event call graph coverage}. Given a GUI event sequence $S = \{e_1, e_2, ..., e_n\}$, we define the call graph coverage of event $e_i$, with $0<i\leq n$ as the ratio between the number of source code statements covered by $S$ to the total number of statements from the methods in $e_i$'s event call subgraph.
\end{definition-non}

For each event, the call graph coverage tells us how many code statements were run out of the maximum possible as determined via static analysis. As defined above, an event's code coverage can be improved if subsequent events run code from its event call graph.

However, all computed call graphs are only approximations. When computed dynamically, that is by running the target application, they might miss methods that can be called via different code paths, and are thus incomplete. When computed statically, using tools such as Soot, they might include additional edges that a more precise analysis might determine to be superfluous. This is a well-known problem in pointer analysis \cite{19}, one that is not expected to be solved for complex languages such as Java. For GUITracer, our approach was to use one of the algorithms within the Soot framework that computes an accurate call graph statically \cite{7}. This means that while all methods that may be called are included in the call graph, it is possible that it also contains superfluous entries.

In order to build event call graphs, we required information about the event handlers installed by the target application. This was achieved using the JETracer framework that is detailed below.

\subsection{The JETRacer Framework}
JETracer is our open-source framework for real-time tracing of GUI events fired within Java applications built using one of the AWT, Swing or SWT toolkits. Available at \cite{16}, the tool consists of two modules: an \emph{agent} and a \emph{host} \cite{10}. The \emph{agent} module is GUI toolkit specific and is deployed within the target application which it instruments during start up. The process is completely transparent to the target application, and as shown within the evaluation section in \cite{10}, it does not impact its perceived performance. The JETracer agent gathers GUI event information and transmits it to the \emph{host} module via network socket. The \emph{host} module maintains the connection with the agent, and once an event is received it notifies any attached listeners. Adding support for new GUI toolkits or events (e.g. touch interactions) is possible by implementing a new \emph{agent} component \cite{10}.

GUITracer is built on top of our JETracer framework, which it uses to gather event information from the target application. Figure \ref{fig:GUITracerArch} shows the architecture of our tool, including the JETracer agent and host. Code coverage capabilities are provided via the open-source JaCoCo\footnote{JaCoCo - http://www.eclemma.org/jacoco/} library that provides coverage information on-demand, during program execution. We harnessed this feature so that every time JETRacer sends event information it also provides updated code coverage via JaCoCo. This is done once for each GUI event fired, when handling routines have completed and control is returned to the GUI toolkit.

\section{The GUITracer Tool}
The GUITracer \cite{21} tool is open-sourced under the Eclipse Public License and is free to download from our website \cite{16}. Fully implemented in Java, the application's only requirement is a Java 6 compatible platform. As a proof of concept implementation, GUITracer comes in the form of a standalone desktop application. A brief video that showcases the tool's main features is available at \cite{21}. In order to set up an application for tracing by our tool is done by providing the location of the following artefacts using command line parameters: 

\begin{itemize}
\item \emph{Source code} - This folder is required in order to enable illustrating source code coverage.
\item \emph{Binaries} - The location of the binaries is required for calculating the application call graph and pre-run instrumentation.
\item \emph{Libraries} - Libraries must be provided separately in order to exclude them from the displayed callgraphs.
\item \emph{Main class} - Required to start the target application.
\item \emph{Call graph} - An optional parameter. As call graph calculation is computationally expensive, storing the call graph between runs decreases the time required to start GUITracer. Our website \cite{16} details how to save generated call graphs in order to be reused on subsequent runs.
\end{itemize}

\begin{figure}
	\centering
	\includegraphics[width=9cm]{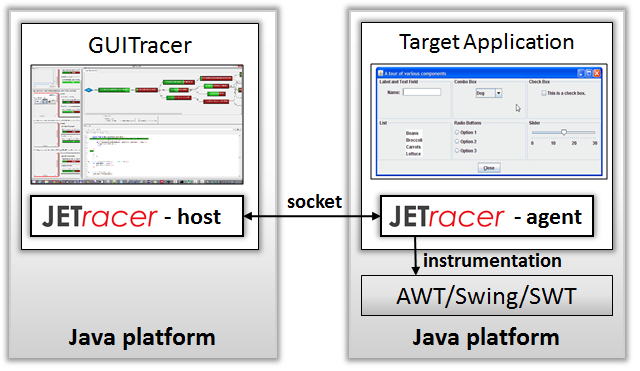}
	\caption{GUITracer tool architecture}
\label{fig:GUITracerArch}
\end{figure}

Figure \ref{fig:GUITracerArch} illustrates GUITracer's architecture. The tool employs our JETracer \cite{10} framework for event and coverage tracing information. Once a GUI event is fired, the JETracer agent relays the information to the framework's host component, which is integrated with GUITracer's designated handler.

Figure \ref{fig:GUITracer} illustrates a screenshot of the application. The target application is version 0.7.1 of the open-source FreeMind\footnote{FreeMind home - http://freemind.sourceforge.net} mind mapping software. The tool's UI consists of three main panes - the \emph{Event trace}, the \emph{Call graph} and the \emph{Source code}. We describe each of them below.

\subsection{Event Trace Pane}
GUI events fired within the target application are displayed within this pane, on the left hand side of Figure \ref{fig:GUITracer}, which currently shows three events. The topmost event is shown with a white background as it is currently selected. The event is an \emph{ActionEvent} fired by the toolbar button having the black outline. For each event, the tool currently displays the following information:

\begin{itemize}
\item \emph{Application window screenshot}. Recorded at the time the event is fired, it allows visually identifying the event's originating widget using its black border outline. It also ensures that any visual styles active within the target application also appear in GUITracer.

\item \emph{Code coverage information}. This provides information regarding the event's code coverage. The colour coding is used consistently with most other coverage tools: green for covered and red for uncovered code, respectively. In addition, as many events might run the same code over and again, we use a lighter shade of green to illustrate code that was first run during the event's handling. 
\end{itemize}

The \emph{application line coverage} section provides information regarding how many code lines have coverage up until and including the event. This is the information most other tools also display, with the difference that in the case of GUITracer, it is gathered after each GUI event is handled. The \emph{event line coverage} section is based on the event coverage concepts introduced within the previous section. It uses the statically computed call graph of the application and shows the event's call graph coverage. Its purpose is to facilitate feature localization as well as provide detail regarding the link between GUI events and the source code that handles them.

As an example, after the topmost event in the trace was handled, target application coverage was of 2082 lines out of a total of 7615. The event call graph comprises 404 code lines, with a coverage of 276 lines. As shown by the light green shade, most statements handling this \emph{ActionEvent} were not run before.

The type of GUI events displayed in the trace can be filtered. This functionality was added due to observing that in many cases, GUI applications fire a large number of events which  clutter the event trace. These include focus events fired when the target application loses/gains focus as well as mouse movement events. In addition to filtering these, users can choose to hide events that do not contribute to the target application's code coverage. This is useful to hide repeated events that always take the same code path. The topmost event in the trace is selected, as shown by the white background. Once this happens, the call graph pane becomes relevant.

\subsection{Call Graph Pane}
\begin{figure*}
	\centering
	\includegraphics[width=\textwidth]{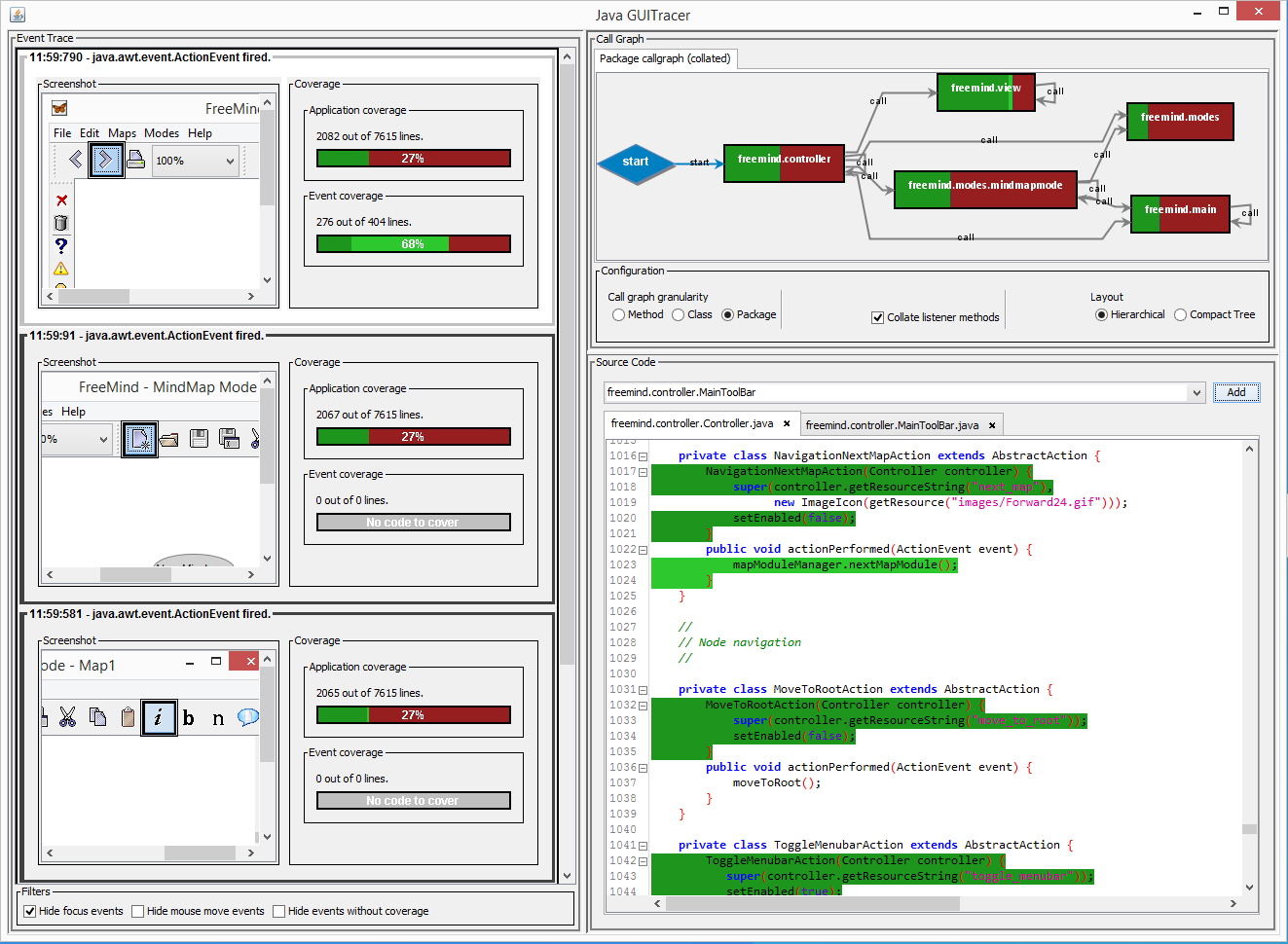}
	\caption{GUITracer screenshot}
\label{fig:GUITracer}
\end{figure*}

Once an event from the trace is selected, GUITracer calculates its event call graph, and displays it in the top pane on the right hand side of Figure \ref{fig:GUITracer}. Each call graph has exactly one \emph{start} node, with one outgoing edge for each handler. The displayed call graph only contains application code; library as well as Java platform calls are not included, and neither are any callbacks from them. This is due to the difficulty of modelling library callbacks, which is an active topic of research \cite{18}. 

The displayed call graph can be customized using the controls  below the call graph panel. First of all, the graph can be displayed with method, class or package granularity. Most detailed call graphs are at method level, where each vertex represents one method. At class and package level, each vertex represents a class or package, respectively. Regardless of call graph granularity, code coverage is shown using consistent colour coding with the trace view. At class or package granularity, method calls are displayed as edges between the classes or packages they belong to, as is the case. 

Changing the granularity changes both the level of detail as well as the complexity of the displayed graph. While this is application specific, method call graphs can easily contain hundreds of vertices, while class and package call graphs usually contain no more than a few dozen. In addition, users can choose to display one collated call graph, which represents all event handlers in the same pane, or have a separate tab for each handler. 

These features are meant to facilitate exploration and feature location, as users can quickly retrieve information regarding the coverage of each source code entity. If a method or class-level call graph is displayed, nodes present a contextual menu allowing the element's source code to be shown in the corresponding panel.

\subsection{Source Code Pane}
This pane allows users to consult the target application's source code. Source files can be opened using the combo-box control or from the contextual menu of the call graph nodes. Source files are displayed with syntax as well as coverage highlighting. The employed colour scheme is consistent with the one previously described. The purpose of the source code pane is to illustrate that by following a top-down exploration strategy, users can start from the target application's GUI and reach the covered statements in its source code. 

Given the technical challenges of implementing the GUITracer tool, this proof of concept was implemented as a standalone application. However, as such tools are more useful when integrated within an IDE, the next version of the tool will be implemented as a plugin for a popular Java IDE, such as Eclipse or Netbeans. As IDE's have advanced source code editing components, the GUITracer plugin will employ them in order to showcase GUI event coverage, similar with how most unit test plugins currently work. 

\section{Related Work}
An important body of work which GUITracer employs is the Soot framework \cite{4}. First detailed in \cite{5}, Soot provides static analysis functions for Java programs, among which several algorithms for obtaining the static call graph \cite{6}. Our tool uses the SPARK algorithm detailed in \cite{7}, a context insensitive algorithm that provides an adequate speed to accuracy trade-off. One of the first tools to employ Soot was JAnalyzer \cite{8}, which leveraged call graph information to provide simple graphical representation of the call relations in the target application. Our tool builds on JAnalyzer by providing complete call subgraphs of the application code starting from the entry points into event handler code.

Our tool's code coverage functionalities are inspired by efforts such as detailed by Duck et al. in \cite{2}, where a new approach for software reconnaissance based on differential code coverage is proposed. The approach is then investigated within an evaluation where users had to debug and change code in several complex GUI-driven applications. Our implementation complements the one detailed in \cite{2} by providing context to coverage information in the form of the GUI event trace, which facilitates identifying the relation between the source code and the user interface controls. An effort more related to GUI-driven systems is detailed in \cite{1}, where authors describe a navigation mechanism that enables source code localization for GUI elements. A controlled user study is also detailed within \cite{1} showing important speed-ups in feature localization tasks.

One of the key challenges of implementing our tool was accurately capturing GUI event information. As this is a software tracing task, we studied previous efforts targeting Java, such as the JMonitor library developed by Karaorman and Freeman \cite{9}. JMonitor provides event monitoring for Java by specifying \emph{event patterns} and \emph{event monitors}. Patterns are used to describe interesting events, and monitors act as handlers that are called once the events have taken place. The proposed library provides a generic implementation for lowest-level events such as setting the value of a class field or a method call. Another notable example is JRapture \cite{11}, a tool for capturing and replaying Java program executions by recording interactions between the program itself and the system, using accurately reproduced input sequences. Profiling can then be added to study the application during replay. JETracer differentiates itself from these tools by working on a higher abstraction level and being developed to record GUI events. This allows capturing additional information such as application screenshots as well as event listener information. 

\section{Future Work and Conclusion}
Our aim for this implementation was to provide the proof-of-concept for a tool that may find many uses during software development and maintenance. We believe the current implementation lays down the foundation for a useful tool to assist in program comprehension and feature localization for GUI-driven applications.

To the best of our knowledge, GUITracer is the first tool that successfully combines static and dynamic analyses for program comprehension of GUI-driven applications. Its creation was guided by findings from studies targeting professional developers such as those in \cite{15,3}, which underline the fact that software exploration and comprehension are most often started at GUI level. We have also taken into account the findings within Storey et al.'s survey of software exploration tools \cite{17} that show a lack of tools proposing a top-down approach.

In order to identify possible future improvements to GUITracer, we undertook a preliminary evaluation using various versions of open-source applications such as FreeMind, jEdit and Azureus. This allowed us to discover what features are important to improve the tool's capabilities. At the present time, these include improving the visualization of large call graphs using better filtering and navigation, adding support for multi-thread programs and library callbacks \cite{18}. A more distant issue is to provide support for unit testing by integrating our tool's visualization capabilities with well known frameworks such as JUnit.

Regarding the tool's deployment, the next step is to integrate the tool as a plugin within popular IDE's such as Eclipse and NetBeans, where GUITracer will be available while running GUI applications. At this point we plan to undertake a user-driven evaluation in order to guide further development that will make the tool as useful as possible to practitioners working on large scale GUI applications.

\bibliographystyle{IEEEtran}
\bibliography{biblio}

\end{document}